\documentclass[twocolumn]{aastex62}
\usepackage{CJK}
\usepackage{xfrac}

\graphicspath{{./}{figures/}}

\shorttitle{Gravitational Self-force Errors}
\shortauthors{Zhu and Gnedin}

\def\phiH{\phi^{(H)}}
\def\phiL{\phi^{(L)}}
\def\phiE{\phi^{(E)}}

\begin{document}
\begin{CJK*}{UTF8}{gkai}
\title{Gravitational Self-force Errors of Poisson Solvers on Adaptively Refined Meshes}

\correspondingauthor{Hanjue Zhu (朱涵珏)}
\email{hanjuezhu@uchicago.edu}
\author[0000-0003-0861-0922]{Hanjue Zhu (朱涵珏)}
\affiliation{Department of Astronomy \& Astrophysics; 
The University of Chicago; 
Chicago, IL 60637, USA}

\author{Nickolay Y. Gnedin}
\affiliation{Particle Astrophysics Center; 
Fermi National Accelerator Laboratory;
Batavia, IL 60510, USA}
\affiliation{Kavli Institute for Cosmological Physics;
The University of Chicago;
Chicago, IL 60637, USA}
\affiliation{Department of Astronomy \& Astrophysics; 
The University of Chicago; 
Chicago, IL 60637, USA}

\begin{abstract}
An error in the gravitational force that the source of gravity induces on itself (\emph{a self-force error}) violates both the conservation of linear momentum and the conservation of energy. If such errors are present in a self-gravitating system and are not sufficiently random to average out, the obtained numerical solution will become progressively more unphysical with time: the system will acquire or lose momentum and energy due to numerical effects. In this paper, we demonstrate how self-force errors can arise in the case where self-gravity is solved on an adaptively refined mesh when the refinement is nonuniform. We provide the analytical expression for the self-force error and numerical examples that demonstrate such self-force errors in idealized settings. We also show how these errors can be corrected to an arbitrary order by straightforward addition of correction terms at the refinement boundaries.
\end{abstract}

\keywords{methods: numerical}

\section{Introduction}
\label{sec:intro}

A common numerical problem in astrophysics is determining the gravitational potential given an arbitrary mass distribution. Computationally, there exist two separate approaches for obtaining the potential - one is by directly solving the Poisson equation
\[
    \nabla^2\phi = 4\pi G\rho
\]
and the other is by computing a convolution
\[
    \phi(\vec{x}) = -G \int d^3x^\prime \frac{\rho(\vec{x}^\prime)}{|\vec{x}-\vec{x}^\prime|}
\]
(in the 3D case). The density $\rho$ in these equations can be either represented as a field on some mesh or as a collection of individual point masses (particles) distributed in space.
While mathematically these two approaches are equivalent, computationally they are implemented with very different algorithms. 

One of the commonly used particle simulation methods is the particle-mesh(PM) algorithm \citep{he1988}. The PM method relies on a mesh-based Poisson solver and solves the Poisson equation on a uniform grid. When modeling particles, the first step of the PM method is to deposit particle masses onto a uniform mesh of density values. On a uniform mesh, we can solve the Poisson equation from the density field using a number of methods. The most common methods use the Fourier transform (usually using the ``fast Fourier transform," or FFT, implementation) to compute the convolution of the density field with the Green's function as a product in the Fourier space.

In addition to the Fourier transform, an alternative class of methods for solving the Poisson equation are the iterative relaxation methods. In a relaxation method, the Poisson equation at each mesh point is approximated by a finite difference analog of the Laplace operator, which is thus reduced to a linear algebraic operator acting on the potential. Solving the Poisson equation is then reduced to solving a system of linear algebraic equations, i.e.\ finding the inverse of a very large matrix. This is most commonly achieved by successive iterations, either with a Krylov-type method such as a conjugate gradient or by solving an equivalent diffusion problem (often called a ``relaxation method"). It is also possible to invert the matrix directly, but for state-of-the-art simulations the equation matrix is usually so large that the direct inversion is not practical. However, iterative relaxation methods often converge quite slowly, especially when the grid size is large, because at each iteration a given grid location exchanges information only with neighboring points and the information needs to propagate over the entire computational domain, thus requiring at least as many iterations as there are grid points along one dimension. This is when multigrid methods come to the rescue. Multigrid methods, quite literally, involve using a hierarchy of grids of decreasing spatial resolution. The computational time can be significantly reduced by iterating first on the coarser grids and then propagating the coarse solution down to progressively finer grids, improving the accuracy of the solution by additional iterations at each step.

Methods that rely on the second approach of computing a convolution begin with the straightforward direct summation of contributions from each of the $N$ particles or mesh locations to every other particle or mesh location. Because direct summation scales as $N^2$, this method is practical only for self-gravitating systems of a modest number of gravitating bodies. Direct summation methods can be accelerated to scale as $N\log N$ or even linearly in $N$ by using hierarchical multipole methods such as a tree \citep{Barnes1986} and the fast multipole methods \citep[FMM;][]{gr87,cgr99}. As the name ``multipole methods" suggests, the essence of the two groups of methods lies in using multipole expansion to describe the gravitational potential generated by matter distribution sufficiently far away from the target location. In addition, both of the multipole methods, as well as the direct summation method, can be supplemented with a PM solver for computing a long-range component of the potential, with the short-range part solved with a tree or an FMM. Examples of such a ``mixed" method are the particle-particle particle-mesh (P$^3$M) algorithm \citep{he1988} and the Tree-PM algorithms used in GADGET-2 and the Arepo codes \citep{gadget2,arepo,arepo2}.

Many problems in astrophysics are inherently multiscale. In some regions of the computational domain, high resolution is needed to adequately solve the equations, whereas in other regions, a coarser resolution may suffice. The adaptive mesh refinement (AMR) class of methods \citep[e.g.][]{ BergerOliger84,Popinet03} provides a means for reaching large spatial and temporal dynamical ranges with reduced computational cost. In AMR, the mesh hierarchy consists of meshes at different refinement levels, with higher refinement levels hosting cells of smaller spatial sizes. Highly refined meshes are often restricted to particular regions of interest in the computational domain that occupy a small fraction of the total volume, thus achieving substantial gain in computational efficiency. The refinement procedure can be repeated recursively until the desired level of spatial resolution is reached. Gravity is most commonly solved on AMR meshes (at least in the field of astrophysics) with the first, the Poisson equation approach, using a relaxation solver, although more complicated FFT-based methods have been used as well \citep[e.g.][]{HuangGreengard99,Ricker08,Passy2014}.

In the field of computational galaxy formation, the most common AMR codes are RAMSES \citep{ramses}, Enzo \citep{Bryan2014}, and adaptive refinement tree (ART) \citep{kravtsov99,kravtsov_etal02,rudd_etal08}. Several other codes are used in other branches of computational astrophysics. 
All of these codes only \textit{approximate} the gravitational potential and the force, and so they introduce some finite and quantifiable errors. An error in the gravitational force is not critical per se because few of the other computational components of a complex simulation are exact. A small error in the gravitational force at the location of a test particle causes a proportionately small deviation in the particle orbit, but does not necessarily result in violation of any of the conservation laws. 

In the case of self-gravity, however, an error in the gravitational force that the source of gravity induces on itself (\emph{a self-force error}) is much more serious as it violates both the conservation of linear momentum and the conservation of energy. If such errors are present and are not sufficiently random to average out, they will result in the numerical solution becoming progressively more unphysical with time, acquiring or loosing momentum and energy due to numerical artifacts. 

\section{Exactly Solvable Example}
\label{sec:ese}

An idealized example of the self-force generated in the solution of the Poisson equation by a mesh refinement boundary can be constructed by considering a formally infinite, in practice large enough computational domain with the $x>0$ half being the region of interest (ROI). The ROI is covered by a uniform regular mesh with cells of size $h$. A point gravitational source is located inside the ROI at location $\vec{r}_s$ (i.e.\ $x_s>0$).

We now consider two cases: in the ``high-resolution" case, the $x<0$ half is covered by the grid of the same resolution $h$ (so that the whole computational domain is a uniform regular grid), with the gravitational potential $\phi^{(H)}$ providing an approximation to the exact gravitational potential $\phi^{(E)}\equiv -GM/|\vec{r}-\vec{r}_s|$ inside the ROI with the precision consistent with the numerical method used to solve for $\phiH$. In the ``low-resolution" case, the $x<0$ half of the computational domain is refined to a lower level of refinement, i.e.\ it is covered by a uniform regular grid with the cells of size $2h$. The potential inside the ROI in this case is $\phiL$. 

We adopt a convention where the physical fields (the density and the gravitational potential) are defined at cell centers (``cell data"), although the exact placement of physical fields is not fundamental here. Thus, in our ``high-resolution" case, the density and the potential are sampled at points $(x,y,z)=h(i+\sfrac{1}{2},j+\sfrac{1}{2},k+\sfrac{1}{2})$ for integers $i,j,k$, while in the ``low-resolution" case, the density and the potential are sampled at points $(x,y,z)=2h(i+\sfrac{1}{2},j+\sfrac{1}{2},k+\sfrac{1}{2})$ with $i<0$.

The high-resolution solution $\phiH$ is an approximate solution to the Poisson equation, and deviates from the exact solution $\phiE$ due to the errors of the Poisson solver used. In this paper, however, we ignore these errors and only consider errors due to the refinement boundary at $x=0$, i.e.\ we assume that $\phiH=\phiE$. Any error due to the approximate Poisson solver will be \emph{in addition} to the errors we discuss here. In particular, errors discussed here will not disappear if a more precise higher-order Poisson solver is used.

Both $\phiH$ and $\phiL$ satisfy the Poisson equation inside the ROI ($x>0$ half of our computational domain), and so the difference between them, $u \equiv \phiL-\phiH$, satisfies the Laplace equation in the ROI, albeit not at the $x=0$ boundary,
\begin{equation}
  \nabla^2 u = 0
  \label{eq:lap}
\end{equation}
The nontrivial boundary condition for $u$ is along the $x=0$ plane, and we use $u_0$ to denote the difference along $x=0$:
\[
  u(0,y,z) = u_0(y,z).
\]
We specify $u_0$ below.

The standard approach for solving this Dirichlet problem is to apply 2D Fourier transform to $u$ and $u_0$ in the $(y,z)$ plane,
\begin{eqnarray}
  u_k(x) & = & \int dy\,dz\, u(x,y,z)\, e^{-i(yk_y+zk_z)} ,\nonumber\\
  u_{0,k} & = & \int dy\,dz\, u_0(y,z)\, e^{-i(yk_y+zk_z)} .\nonumber\\
\end{eqnarray}
In this case, the Laplace equation for $u$ becomes an ODE for $u_k$,
\[
  \frac{d^2u_k}{dx^2} -k^2 u_k(x) = 0,
\]
with the solution
\begin{equation}
  u_k(x) = u_{0,k} e^{-kx}.
  \label{eq:uk}
\end{equation}
The solution for $u$ can now be written as a convolution over the boundary condition $u_0$:
\begin{eqnarray}
  u & = & \frac{1}{4\pi^2} \int d^2k\, u_{0,k}\, e^{i(yk_y+zk_z)-kx} \nonumber\\
  & = & \int dy^\prime\,dz^\prime\, G_x(y-y^\prime,z-z^\prime) u_0(y^\prime,z^\prime)
  \label{eq:usol1}
\end{eqnarray}
with the Green's function
\begin{eqnarray}
  G_x(y,z) & = & \frac{1}{4\pi^2} \int d^2k\,  e^{i\left(yk_y+zk_z\right)-kx}\nonumber\\
  & = & \frac{1}{2\pi}\frac{x}{\left(x^2+y^2+z^2\right)^{3/2}}.
  \label{eq:u}
\end{eqnarray}

\begin{figure}[t]
\centering\includegraphics[width=0.9\columnwidth]{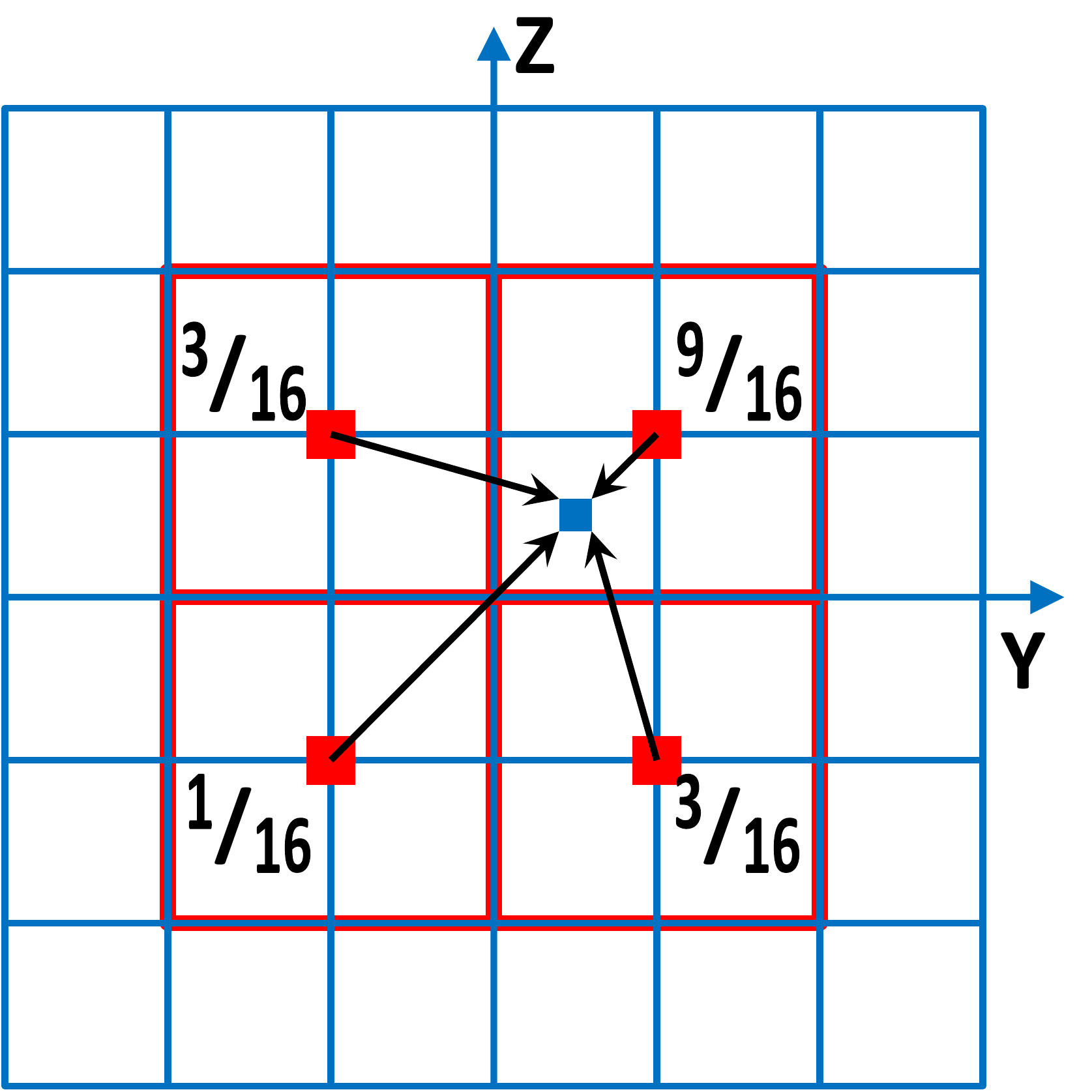}
\caption{Sketch of linear interpolation \citep[called prolongation operator $P_2$ in][]{Guillet2011} from the low-resolution solution $\phiL$ to the high-resolution solution $\phiH$ on the boundary $x=0$.}\label{fig:sketch}
\end{figure}

The boundary condition $u_0(y,z)$ is the difference between the low- and high-resolution solutions on the boundary. If the interpolation is of infinitely high order, then no error is introduced. In practice, however, linear interpolation is commonly used. We adopt the terminology used in \citet{Guillet2011} hereafter. Values of the potential $\phiH$ in the high-resolution mesh at the $x=0$ boundary is the result of the linear prolongation operator $P_2$ acting on the low-resolution mesh that covers the $x=0$ boundary. We show a 2D sketch describing the interpolation in Figure \ref{fig:sketch}. We emphasize that the sketch shows the 2D $x=0$ boundary in the 3D space and shows interpolation on the cell edges. Because we do not consider any specific Poisson solver here, we assume that the Poisson solver uses the boundary values in computing the potential near the boundary. This is indeed the case in the real AMR codes we are familiar with, such as ART and RAMSES.

Ignoring errors of the Poisson solver and assuming that values of the potential on the red and blue squares are exact, the error due to the $P_2$ interpolation at the center of the high-resolution boundary cell $(x,y,z) = (0,h/2,h/2)$ is
\begin{equation}
    u_0(y,z) = \phiL_B\left(\frac{h}{2},\frac{h}{2}\right)-\phiH_B\left(\frac{h}{2},\frac{h}{2}\right),
    \label{eq:boderr}
\end{equation}
where $\phi_B(y,z)$ is the potential at the $x=0$ boundary. We apply the P2 operator to the low-resolution mesh and obtain
\begin{eqnarray}
    \phiL_B\left(\frac{h}{2},\frac{h}{2}\right) & = & \frac{1}{16}\left[\phiL_B(-h,-h)+3\phiL_B(h,-h)+\right.\nonumber\\
    & & \left.3\phiL_B(-h,h)+9\phiL_B(h,h)\right].
    \label{eq:eqLP2}
\end{eqnarray}
Since we ignore errors due to the Poisson solver, we can equate $\phi_B$ with the exact solution $\phiE$, or, equivalently, $\phiH$. Hence, after Taylor expanding $\phi_B$ around $(\sfrac{h}{2},\sfrac{h}{2})$, to the first nonvanishing order in $h$, 
\begin{equation}
    u_0(y,z) = \frac{3}{8}h^2\left(\frac{\partial^2\phiE}{\partial y^2}+\frac{\partial^2\phiE}{\partial z^2}\right).
    \label{eq:u0}
\end{equation}
Because of the symmetry of the problem, this expression is valid not only for $(h/2,h/2)$ but for all $(y,z)$.
Equations (\ref{eq:u}) and (\ref{eq:u0}) give an error in the gravitational potential in the ROI due to the presence of the low-resolution region $x<0$. Note that this error is independent of the actual numerical method used for solving the Poisson equation.

In the case where we place a point source at location $\vec{r}_s$,
\begin{equation}
u_0(y,z) = \frac{3}{8}h^2 GM \frac{2x_{s}^{2}-\left(y-y_{s}\right)^{2}-\left(z-z_{s}\right)^{2}}{\left|\vec{r}-\vec{r}_{s}\right|^{5}}. \label{eq:u0ps}
\end{equation}
Plugging Equation (\ref{eq:u0ps}) into Equation (\ref{eq:usol1}), we obtain
\begin{eqnarray}
u(x,y,z) & = & \frac{3}{16\pi} h^{2} G M \int d y^{\prime} d z^{\prime} \nonumber \\ 
& & \frac{x}{\left[x^{2}+\left(y-y^{\prime}\right)^{2}+(z-z^{\prime})^{2}\right]^{3 / 2}} \nonumber \\ 
& &\frac{2x_{s}^{2}-\left(y^{\prime}-y_{s}\right)^{2}-\left(z^{\prime}-z_{s}\right)^{2}}{\left[x_{s}^{2}+\left(y^{\prime}-y_{s}\right)^{2}+\left(z^{\prime}-z_s\right)^{2}\right]^{5 / 2}} .
\end{eqnarray}

The most significant effect of such an error is that it does not vanish at the location of the point source $\vec{r}_s$, i.e.\ \emph{it induces a self-force}. To calculate the self-force, we can first take $y_{s} = z_{s} = 0$, because of the symmetry of the problem. We are interested in the force at $(x,y,z) = (x_s,y_s,z_s) = (x_s,0,0)$. From the symmetry, it is obvious that
\[
  \frac{\partial u}{\partial y}|_{(x_{s}, y_{s}, z_{s})}=\frac{\partial u}{\partial z}|_{(x_{s}, y_{s}, z_{s})}=0.
\]
Thus, when computing the self-force, we only need to compute $\sfrac{\partial u}{\partial x}|_{(x_{s}, 0, 0)}$. With these simplifications,
\begin{eqnarray}
u(x,0,0) & = & \frac{1}{2 \pi} \frac{3}{8} h^{2} G M \int d y^{\prime} d z^{\prime} \nonumber\\ 
& &\frac{x}{\left(x^{2}+y^{\prime 2}+z^{\prime 2}\right)^{3 / 2}} \frac{2 x_{s}^{2}-\left(y^{\prime 2}+z^{\prime 2}\right)}{\left(x_{s}^{2}+y^{\prime 2}+z^{\prime 2}\right)^{5 / 2}}. \nonumber
\end{eqnarray}
Taking $y^{\prime} = \rho$cos$\theta$, $z^{\prime} = \rho$sin$\theta$, we find
\begin{equation}
u(x,0,0) = \frac{3GM}{8} h^{2}\int_{0}^{\infty} \frac{x\rho d \rho }{\left(x^{2}+\rho ^{2}\right)^{3 / 2}} \frac{2 x_s^{2}-\rho^{2}}{\left(x_{s}^{2}+\rho^{2}\right)^{5 / 2}}.
\label{eq:uofx}
\end{equation}

The self-force is the force that the source exerts on itself, i.e.\ the self-acceleration is
\[
  \vec{g}^{ x}_{\rm SF} = \left.\frac{\partial u}{\partial x}\right|_{\left(x_{s}, 0,0\right)} = -\frac{3GM}{8} h^{2} \int_{0}^{\infty} \rho d \rho \frac{\left(\rho^{2}-2 x_{s}^{2}\right)^{2}}{\left(x_{s}^{2}+\rho^{2}\right)^{5}}
\]
\begin{equation}
    = -\frac{9}{64} \frac{h^{2}GM}{x_{s}^{4}}.
    \label{eq:gsf}
\end{equation} 

Equation (\ref{eq:gsf}) is valid for any Poisson solver that uses the boundary condition (\ref{eq:u0}).

\section{Numerical Example}

To illustrate the effect of the self-force in a numerical example, we solve the Poisson equation in a 2D computational volume (to make this example computationally cheap) of size 1. The $x<0.5$ half of this volume is covered by a uniform grid with cells of size $2h$. The $x>0.5$ half of the volume is covered by a mesh twice as fine, with cell size $h$, and we consider $h=1/512$ and $2h=1/256$.

The Poisson equation is solved on this grid using a simple relaxation solver with the Gauss-Seidel relaxation scheme and successful over-relaxation (SOR) acceleration, implemented exactly as described in ``Numerical Recipes in C++" \citep{nr}, with the interpolation on the refinement boundary $x=0.5$ performed with the $P_2$ operator as described in \S\ \ref{sec:ese} and shown in Fig.\ \ref{fig:sketch}. The external boundary conditions are taken to be periodic to avoid introducing any additional source of gravity (because of symmetries, forces from the periodic images of a source in the computational box cancel out exactly).

In all of the examples shown below, the number of iterations of the relaxation solver is set by the requirement of ``full convergence" - i.e.\ when the numerical results computed with the number of iterations $N$ and the results computed with the number of iterations $3N$ do not differ by more than the thickness of the lines in the figures below. This typically requires $N$ to be between 3,000 and 30,000 for our chosen mesh sizes. This high number of iterations is an overkill for any practical calculation; we use them here to make sure that we compute the self-force accurately even when it is small, and without any additional numerical errors due to, for example, incomplete convergence of the Poisson solver.

\begin{figure}[t]
\centering%
\includegraphics[width=\columnwidth]{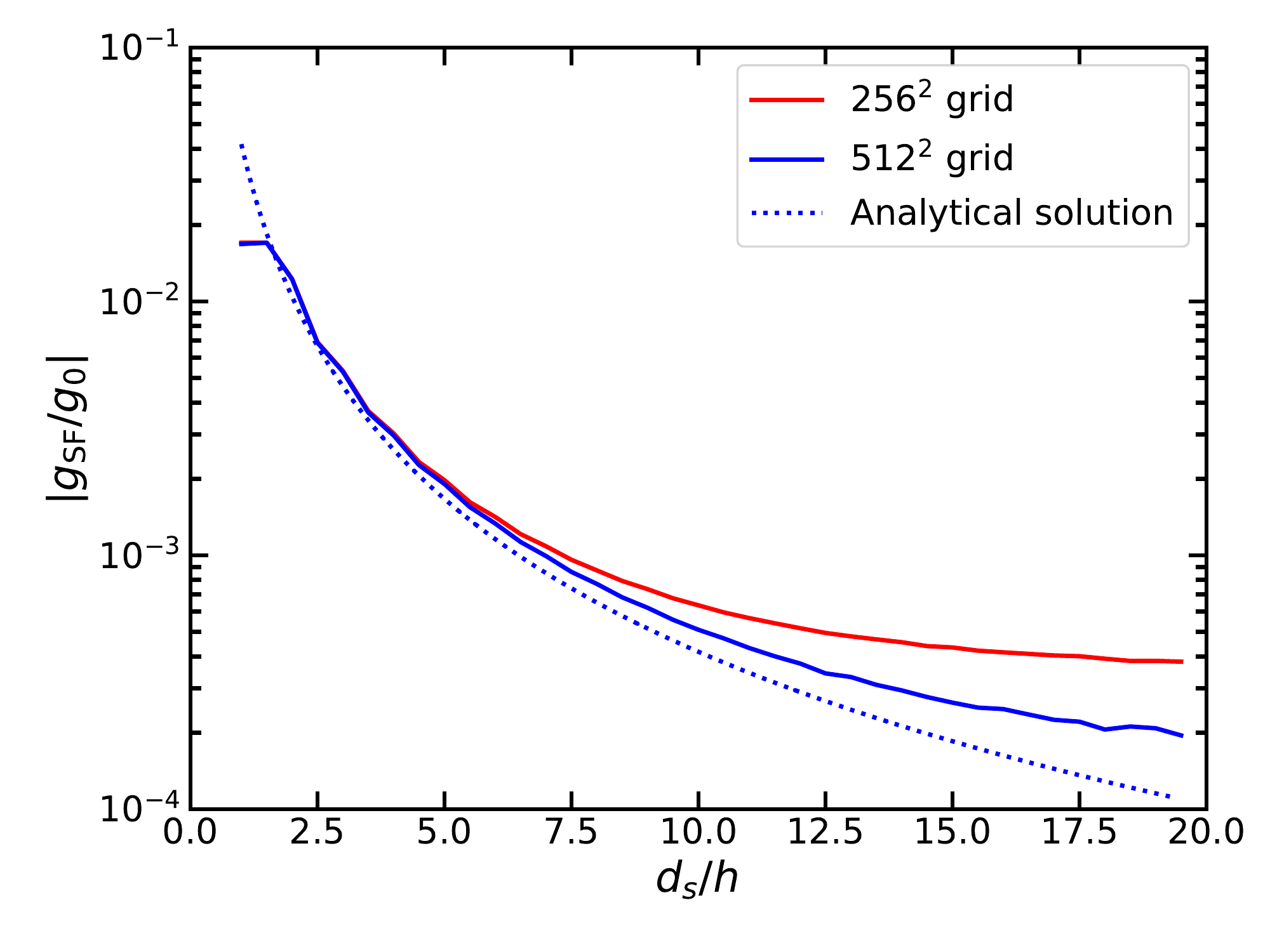}%
\caption{Self-acceleration on a single stationary source of gravity at a distance $d_s$ from the refinement boundary in units of the gravitational acceleration at the boundary, $g_0=GM/d_s$ (we recall that this is a 2D case). The analytical solution matches the numerical results well. We show results with two different grid sizes to illustrate that the discrepancy with the analytical solution at large $d_s$ is due to the periodic boundary conditions in a finite computational volume.}\label{fig:acc}
\end{figure}

In the first test, to compare our analytical calculation with the numerical one, we place a single stationary source of gravity at $(x,y)=(0.5+d_s,0.5)$ (i.e.\ $d_s$ is defined as the distance away from the boundary in the x-direction) and assign the mass density on the grid by the standard clouds-in-cell (CIC) deposition \citep{he1988}. Placement of the source in this and the subsequent tests is such that the CIC assignment is done entirely within the high-resolution region, so there are no additional numerical errors introduced due to the CIC assignment straddling the refinement boundary.

Results of these calculations are shown in Figure \ref{fig:acc}. The numerical self-acceleration matches the analytical calculation (Eq.\ \ref{eq:sf2d} below) very well, except when $d_s$ is large (in other words, when the source of gravity is at a large distance away from the refinement boundary). In this limit, the self-force falls off so much that the effect of our adopted periodic boundary conditions in a finite computational volume becomes noticeable.

\begin{figure}[t]
\centering%
\includegraphics[width=\columnwidth]{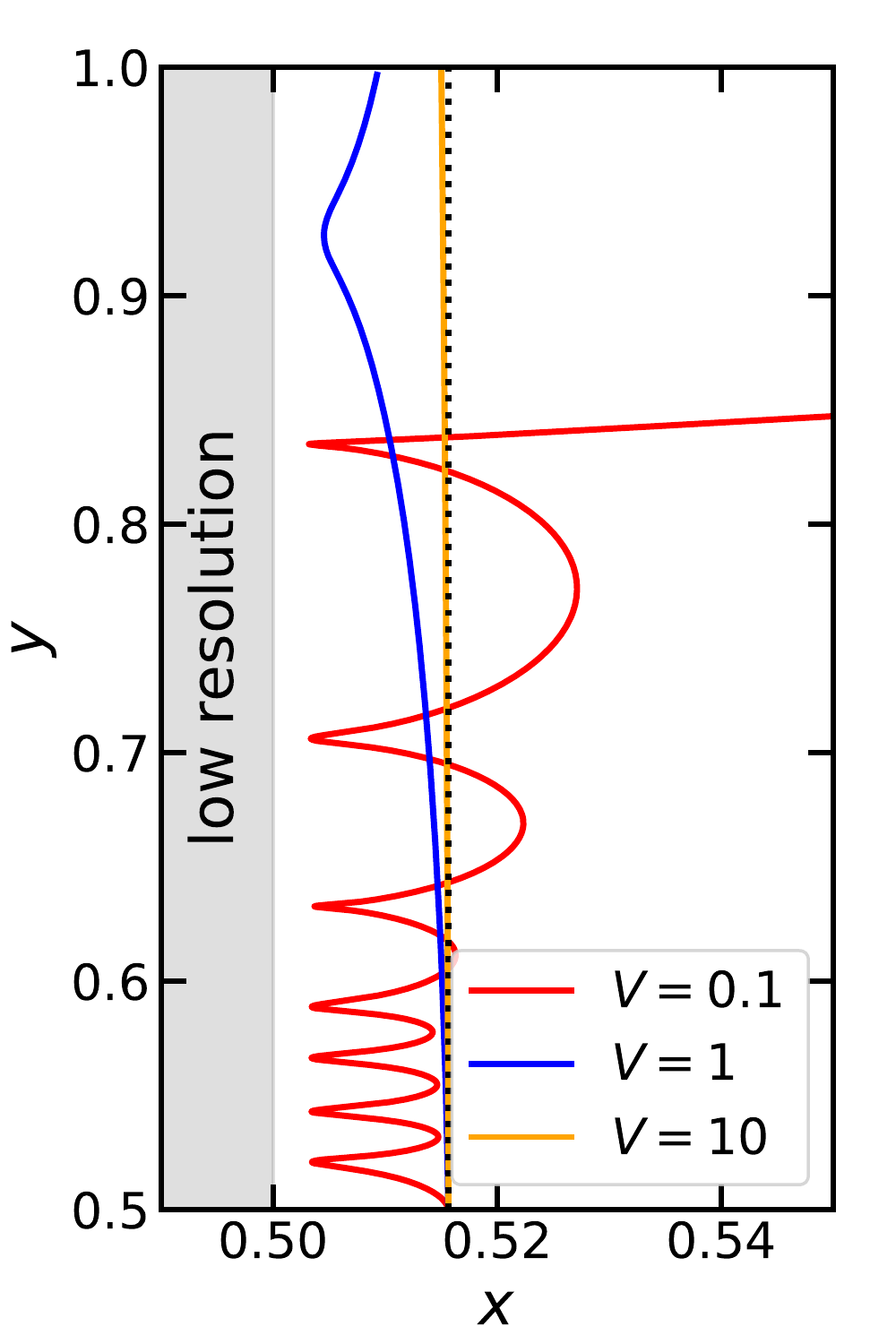}%
\caption{Single particle tests with different particle velocities. Initial particle velocities are in the $y$-direction. Units are such that a particle with velocity $v=1$ crosses the computational box in unit time. A particle starts at the bottom of the figure and proceeds vertically up. In absence of the self-force, it would proceed along the dotted black line. The self-force from Equation (\ref{eq:gsf}) causes the particle to approach the boundary; close to the boundary, the self-force changes sign and the particle is bounced back, gradually acquiring erroneous momentum and energy. The magnitude of the cumulative effect of the self-force is a strong function of particle velocity. Note that the aspect ratio of the figure is not 1.}\label{fig:one}
\end{figure}

In the second test, we examine trajectories of a single particle, initially placed at $(x,y)=(0.5+4h,0.5)$ and given the initial velocity $(\dot{x},\dot{y})=(0,v)$ with different values of $v$. Particle trajectories are integrated in time using the conventional leapfrog integrator with a constant time-step to ensure the conservation of energy. We show the trajectories in Figure \ref{fig:one}. At large $d_s \gg h$, Equation (\ref{eq:gsf}) is valid, and the self-force causes the particle to approach the boundary. At $d_s<1.5h$, the self-force changes sign and the particle is bounced back. Because the self-force is a numerical error, it is not in general a conservative force. Hence, the particle gradually acquires erroneous momentum and energy.

This particular test is not illuminating, however, because the self-force is a function of location only. The relative effect of the self-force on the particle trajectory is therefore dependent on the particle velocity: increasing the particle velocity makes the effect smaller, while the effect can be arbitrarily large for an arbitrarily slow-moving particle. Because for a single source of gravity there is no characteristic velocity, the relative effect of the self-force cannot be demonstrated by this test.

\begin{figure*}[t]
\centering%
\includegraphics[width=0.429\textwidth]{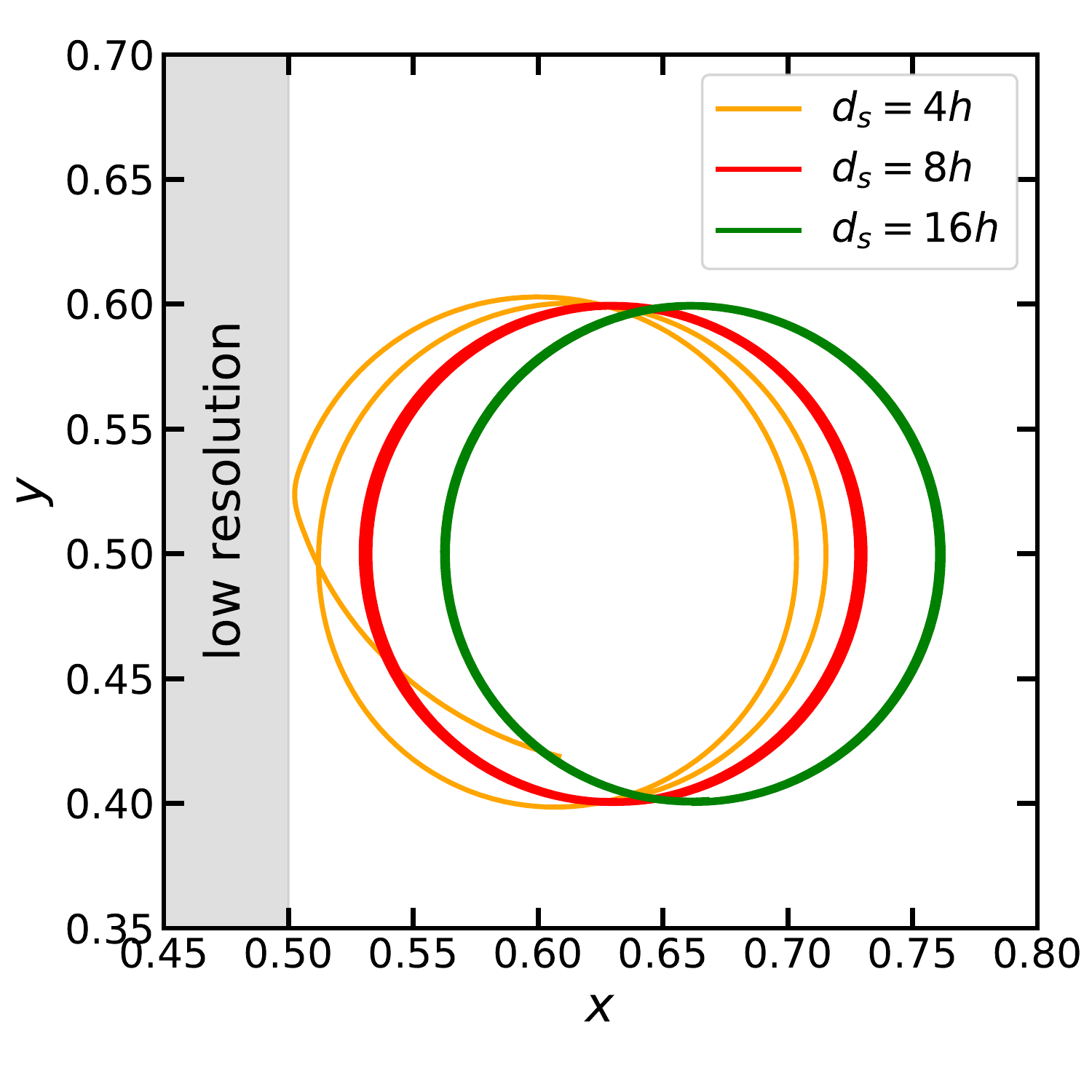}%
\includegraphics[width=0.571\textwidth]{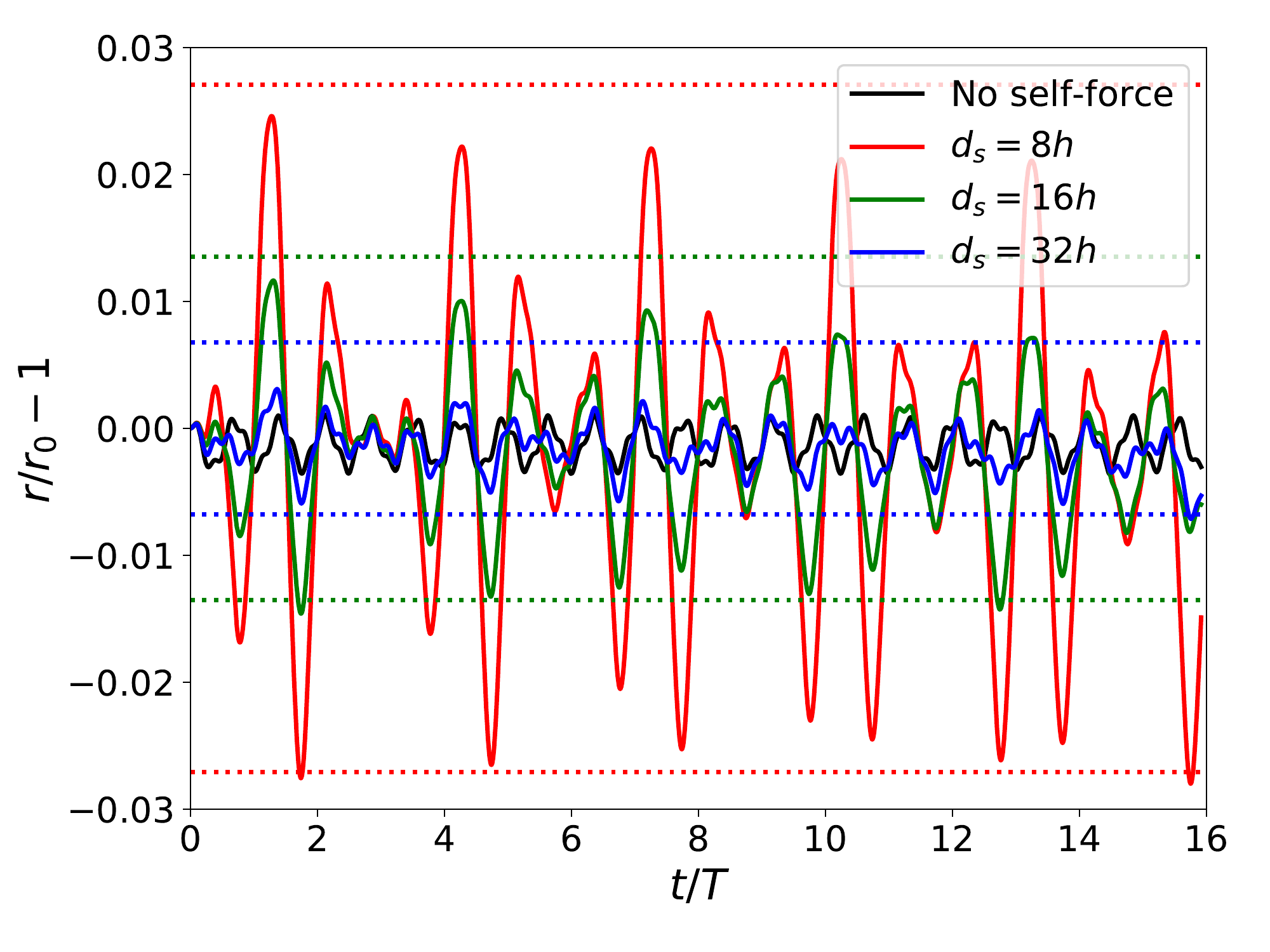}%
\caption{Example of numerical errors introduced by the self-force due to the refinement boundary for two equal-mass particles in a circular orbit in 2D. The left panel shows orbits of one of the particles for different values of $d_s$, the distance of closest approach to the boundary. The closest orbit at $d_s=4h$ is disrupted and is only plotted until the moment soon after the particle orbit suffers a catastrophic error at the refinement boundary. The right panel shows the deviation of orbits from the analytically expected perfect circle $r=r_0$ as a function of time in units of the orbit period. The black line in the right panel also shows a case with gravity computed on a uniform grid covering the entire box with a Particle-Mesh method; in this case, there is no error due to refinement, so the black line shows the accuracy of integrating particle orbits and of the density assignment on the grid. These errors are much smaller than the errors introduced by the refinement boundary. Horizontal dotted lines are analytical estimates for the induced errors from Equation (\ref{eq:maxerr}).\label{fig:numex}}
\end{figure*}

For a more interesting dynamic test, we place two equal-mass particles in a circular orbit of radius $r_0$ along the $y$-axis at distance $d_B > r_0$ from the refinement boundary (i.e.\ at $(x,y)=(0.5+d_B,0.5+r_0)$ and $(0.5+d_B,0.5-r_0)$, respectively. The leapfrog integrator is used in this test as well. In this case the velocities of the two particles are related to the gravitational acceleration between them, and hence the effect of the self-force is only dependent on the distance of the particle closest approach to the refinement boundary.

Orbits of one of the two particles and the corresponding deviations from the ideal circular orbit are shown in Figure \ref{fig:numex} for several values of $d_s= d_B-r_0$, the distance of closest approach to the boundary. The numerical solution is iterated to convergence (numerical errors are below the line thickness) at each time step to ensure that self-force errors are not confused with other numerical effects. 

The error in particle orbits introduced by the self-force can be calculated analytically in the limit of a small self-force. The effective potential of a particle in a circular orbit in 2D is
\begin{equation}
    \psi(r) = \frac{L^2}{2 M r^2} + GM \ln\left(\frac{r}{r_*}\right),
\end{equation}
where $L$ is the angular momentum and $r_*$ is an arbitrary value. For a circular orbit at radius $r_0$, $L^2=G M^2 r_0^2$ and the energy of the orbit is $E_0 = \psi(r_0) = GM/2+GM\ln(r_0/r_*)$.

In the presence of the external acceleration $g_{\rm SF}$ (and the corresponding potential $\phi_{\rm SF}$), the particle at radius $r$ acquires kinetic energy $K$, so that
\begin{equation}
    E_0 = K + \psi(r)+\phi_{\rm SF}(\vec{r}).
    \label{eq:k}
\end{equation}

\begin{figure*}[t]
\centering%
\includegraphics[width=0.429\textwidth]{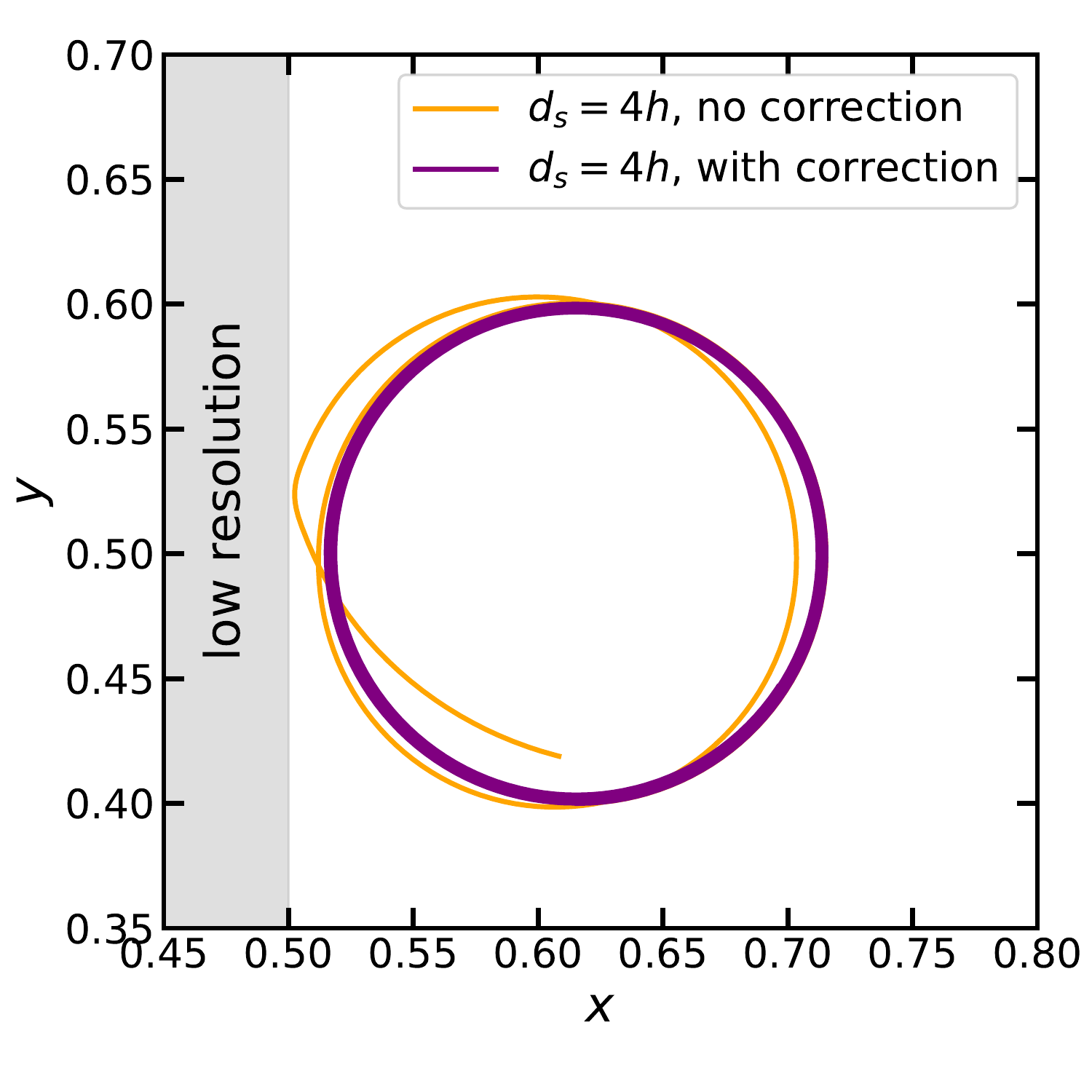}%
\includegraphics[width=0.571\textwidth]{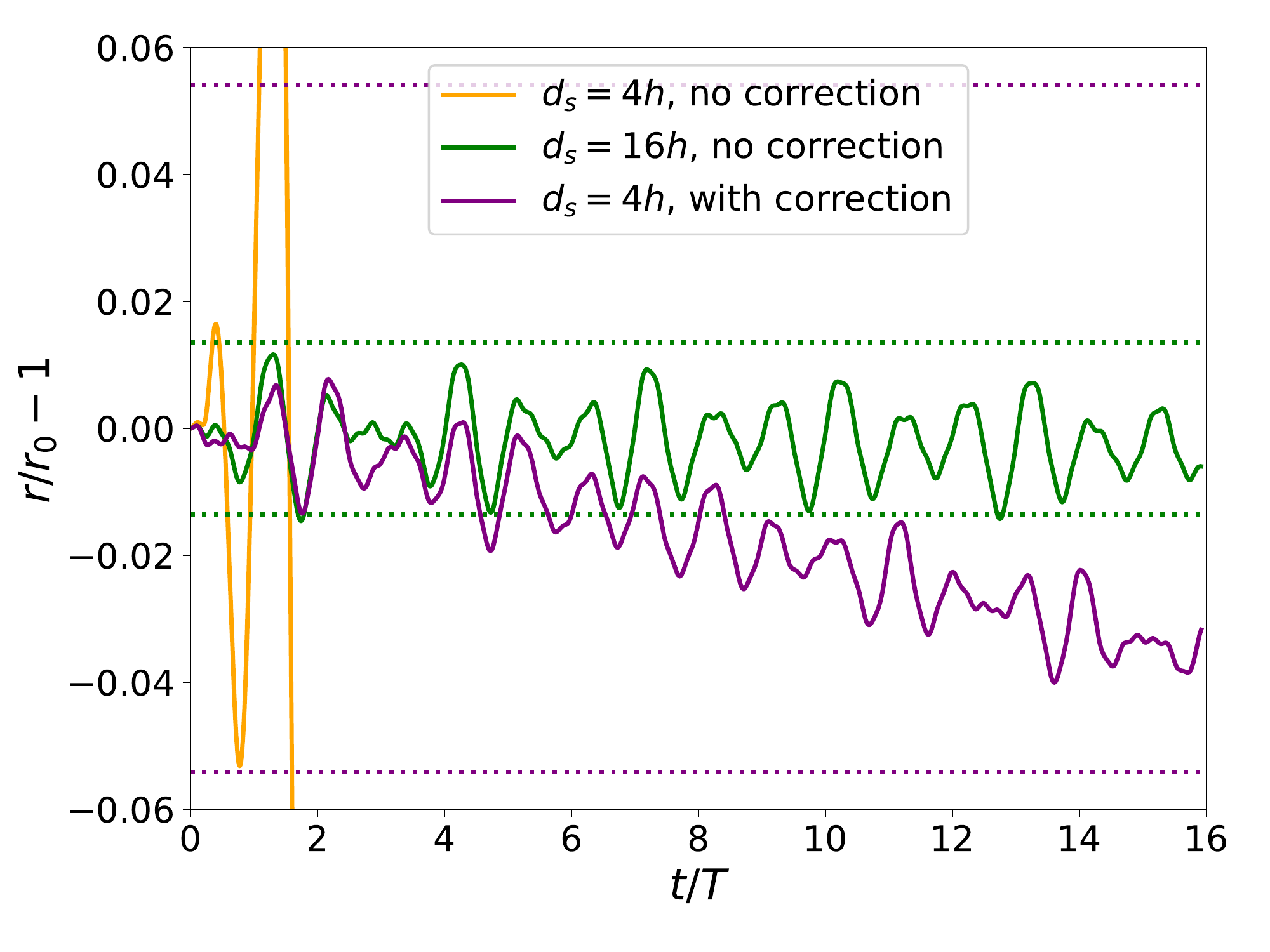}%
\caption{An analog of Fig.\ \ref{fig:numex}, now showing a numerical solution with the boundary correction from Equation (\ref{eq:corr}) .\label{fig:corr}}
\end{figure*}

In the limit where the external acceleration is small, we can consider only small fluctuations around the circular orbit. Taking $\vec{r}=\vec{r}_0(t) + r_0\vec{\xi}$ and considering only perturbations parallel to the radial direction $\vec{r}_0$ that distort the orbit (perturbations along the orbit only affect the phase of the particle and are not visible in Fig.\ \ref{fig:numex}), we find
\[
K = -\phi_{\rm SF}(\vec{r}_0) - GM\xi_\parallel^2.
\]
In our 2D case,
\begin{equation}
  \phi_{\rm SF}(r_0) = -\frac{3h^2GM}{64 (x-0.5)^2}.
\label{eq:sf2d}
\end{equation}

The range of the fluctuations around the ideal circular orbit is set by the condition $K=0$. Hence, the maximum error in the particle position is
\begin{equation}
  \max\left(\frac{r}{r_0}-1\right) = \max(\xi_\parallel) = \frac{\sqrt{3}}{8} \frac{h}{d_s}.
  \label{eq:maxerr}
\end{equation}
This error estimate is shown in Fig.\ \ref{fig:numex} with horizontal dotted lines. For high values of $d_s$ the analytical result matches the numerical result very well.

\section{Correcting the Self-force Error}

The self-force error is generated by the interpolation error on the boundary from Equation (\ref{eq:u0}). The error can be corrected either by using a higher-order interpolation on the boundary, or by including the actual self-force error with a negative sign as a correction. The correction can be made by simply replacing the term $\phiL_B(h/2,h/2)$ from Equation (\ref{eq:boderr}) with
\begin{equation}
    \tilde{\phi}^{(L)}_B = \phiL_B - \frac{3}{8}h^2\left(\frac{\partial^2\phiL_B}{\partial y^2}+\frac{\partial^2\phiL_B}{\partial z^2}\right).
    \label{eq:corr}
\end{equation}
For a simple second-order finite difference approximation to the Laplacian, 
\[
    \frac{\partial^2\phiL_B}{\partial y^2} +\frac{\partial^2\phiL_B}{\partial z^2} \approx \frac{1}{4h^2}\left[\phiL_B(-h,h) + \phiL_B(3h,h) + \right.
\]
\[
     \left.\phiL_B(h,-h) + \phiL_B(h,3h) - 4\phiL_B(h,h)\right],
\]
equation (\ref{eq:corr}) becomes
\[
    \tilde{\phi}^{(L)}_B \approx \frac{1}{32} \left[30 \phiL_B(h,h)+2 \phiL_B(-h,-h)+3 \phiL_B(-h,h)+\right.
\]
\[
    \left.3 \phiL_B(h,-h)-3 \phiL_B(h,3 h)-3 \phiL_B(3h,h)\right],
\]
which is simply one of the several possible forms of the second-order interpolation on the boundary (resulting in the self-force being reduced to the third order). Hence, using the higher-order interpolation on the boundary is equivalent to using correction (\ref{eq:corr}). This equivalence is not surprising because the Poisson equation is linear.

The result of applying the correction (\ref{eq:corr}) for the two orbiting particle test is shown in Figure \ref{fig:corr} in the extreme case of $d_s=4h$. The orbit is now much more stable, even though it approaches the refinement boundary to within four cells. We note that the amplitude of the fluctuations around the exact solution in the beginning is comparable to the case with $d_s=16h$, which is consistent with the order of $u_0$ from Equation (\ref{eq:u0}) and Equation (\ref{eq:maxerr}). The higher-order terms in both equations eventually contribute to the self-force error, and the orbit moves away from the refinement boundary as indicated by the downward trend of the purple line. This higher-order error can also be corrected by further expanding $u_0$ in powers of $h/d_s$. 

In summary, while refinement boundaries do introduce self-force errors when gravity is computed with a Poisson solver, such as a relaxation solver, these errors are limited in range (fall off as $(h/d)^2$ with $d$ being the distance from the boundary) and can easily be corrected to a higher order in $h/d$ by including simple correction terms at the refinement boundaries.

\acknowledgements
We thank Andrey Kravtsov and Romain Teyssier for valuable suggestions that corrected mistakes in earlier version of this paper and the anonymous referee for the constructive comments that improved the original manuscript. This document was prepared using the resources of the Fermi National Accelerator Laboratory (Fermilab), a U.S. Department of Energy, Office of Science, HEP User Facility. Fermilab is managed by Fermi Research Alliance, LLC (FRA), acting under Contract No. DE-AC02-07CH11359.

\bibliography{main}

\begin{thebibliography}{}
\expandafter\ifx\csname natexlab\endcsname\relax\def\natexlab#1{#1}\fi
\providecommand{\url}[1]{\href{#1}{#1}}

\bibitem[{{Barnes} \& {Hut}(1986)}]{Barnes1986}
{Barnes}, J., \& {Hut}, P. 1986, \nat, 324, 446

\bibitem[{{Berger} \& {Oliger}(1984)}]{BergerOliger84}
{Berger}, M.~J., \& {Oliger}, J. 1984, Journal of Computational Physics, 53,
  484

\bibitem[{{Bryan} {et~al.}(2014){Bryan}, {Norman}, {O'Shea}, {Abel}, {Wise},
  {Turk}, {Reynolds}, {Collins}, {Wang}, {Skillman}, {Smith}, {Harkness},
  {Bordner}, {Kim}, {Kuhlen}, {Xu}, {Goldbaum}, {Hummels}, {Kritsuk}, {Tasker},
  {Skory}, {Simpson}, {Hahn}, {Oishi}, {So}, {Zhao}, {Cen}, {Li}, \& {Enzo
  Collaboration}}]{Bryan2014}
{Bryan}, G.~L., {Norman}, M.~L., {O'Shea}, B.~W., {et~al.} 2014, \apjs, 211, 19

\bibitem[{Cheng {et~al.}(1999)Cheng, Greengard, \& Rokhlin}]{cgr99}
Cheng, H., Greengard, L., \& Rokhlin, V. 1999, Journal of Computational
  Physics, 155, 468 .
\newblock
  \url{http://www.sciencedirect.com/science/article/pii/S0021999199963556}

\bibitem[{Greengard \& Rokhlin(1987)}]{gr87}
Greengard, L., \& Rokhlin, V. 1987, Journal of Computational Physics, 73, 325 .
\newblock
  \url{http://www.sciencedirect.com/science/article/pii/0021999187901409}

\bibitem[{{Guillet} \& {Teyssier}(2011)}]{Guillet2011}
{Guillet}, T., \& {Teyssier}, R. 2011, Journal of Computational Physics, 230,
  4756

\bibitem[{{Hockney} \& {Eastwood}(1988)}]{he1988}
{Hockney}, R.~W., \& {Eastwood}, J.~W. 1988, {Computer simulation using
  particles}

\bibitem[{Huang \& Greengard(1999)}]{HuangGreengard99}
Huang, J., \& Greengard, L. 1999, SIAM Journal of Scientific Computing, 21,
  1551

\bibitem[{{Kravtsov}(1999)}]{kravtsov99}
{Kravtsov}, A.~V. 1999, PhD thesis, NEW MEXICO STATE UNIVERSITY

\bibitem[{{Kravtsov} {et~al.}(2002){Kravtsov}, {Klypin}, \&
  {Hoffman}}]{kravtsov_etal02}
{Kravtsov}, A.~V., {Klypin}, A., \& {Hoffman}, Y. 2002, \apj, 571, 563

\bibitem[{{Passy} \& {Bryan}(2014)}]{Passy2014}
{Passy}, J.-C., \& {Bryan}, G.~L. 2014, \apjs, 215, 8

\bibitem[{{Popinet}(2003)}]{Popinet03}
{Popinet}, S. 2003, Journal of Computational Physics, 190, 572

\bibitem[{{Press} {et~al.}(2002){Press}, {Teukolsky}, {Vetterling}, \&
  {Flannery}}]{nr}
{Press}, W.~H., {Teukolsky}, S.~A., {Vetterling}, W.~T., \& {Flannery}, B.~P.
  2002, {Numerical recipes in C++ : the art of scientific computing}

\bibitem[{{Ricker}(2008)}]{Ricker08}
{Ricker}, P.~M. 2008, \apjs, 176, 293

\bibitem[{{Rudd} {et~al.}(2008){Rudd}, {Zentner}, \& {Kravtsov}}]{rudd_etal08}
{Rudd}, D.~H., {Zentner}, A.~R., \& {Kravtsov}, A.~V. 2008, \apj, 672, 19

\bibitem[{{Springel}(2005)}]{gadget2}
{Springel}, V. 2005, \mnras, 364, 1105

\bibitem[{{Springel}(2010)}]{arepo}
---. 2010, \mnras, 401, 791

\bibitem[{{Teyssier}(2002)}]{ramses}
{Teyssier}, R. 2002, \aap, 385, 337

\bibitem[{{Weinberger} {et~al.}(2020){Weinberger}, {Springel}, \&
  {Pakmor}}]{arepo2}
{Weinberger}, R., {Springel}, V., \& {Pakmor}, R. 2020, \apjs, 248, 32

\end{thebibliography}

\end{CJK*}
\end{document}